\documentclass[english,aip,jcp,amssymb,reprint]{revtex4-1}
\usepackage{multirow}
\usepackage{graphicx}
\usepackage{babel}
\usepackage{amsmath}
\usepackage{enumerate}
\def\vec#1{\textbf{\textit{#1}}}

\begin{document}

\title{Atomic scale friction of molecular adsorbates during diffusion}

\author{B. A. J. Lechner}
\email{bajl2@cam.ac.uk}
\thanks{BAJL and ASdW contributed equally to this work}
\affiliation{Cavendish Laboratory, University of Cambridge, JJ Thomson Avenue, Cambridge, CB3 0HE, United Kingdom}

\author{A. S. de Wijn}
\email{dewijn@fysik.su.se}
\thanks{BAJL and ASdW contributed equally to this work}
\affiliation{Department of Physics, Stockholm University, 106 91 Stockholm, Sweden}

\author{H. Hedgeland}
\author{A. P. Jardine}
\affiliation{Cavendish Laboratory, University of Cambridge, JJ Thomson Avenue, Cambridge, CB3 0HE, United Kingdom}
\author{B. J. Hinch}
\affiliation{Department of Chemistry and Chemical Biology, Rutgers University, Piscataway, New Jersey 08854, USA}
\author{W. Allison}
\author{J. Ellis}
\affiliation{Cavendish Laboratory, University of Cambridge, JJ Thomson Avenue, Cambridge, CB3 0HE, United Kingdom}

\date{\today}

\begin{abstract}
Experimental observations suggest that molecular adsorbates exhibit a larger friction coefficient than atomic species of comparable mass, yet the origin of this increased friction is not well understood. We present a study of the microscopic origins of friction experienced by molecular adsorbates during surface diffusion. Helium spin-echo measurements of a range of five-membered aromatic molecules, cyclopentadienyl (Cp), pyrrole and thiophene, on a copper(111) surface are compared with molecular dynamics simulations of the respective systems. The adsorbates have different chemical interactions with the surface and differ in bonding geometry, yet the measurements show that the friction is greater than 2~ps$^{-1}$ for all these molecules. We demonstrate that the internal and external degrees of freedom of these adsorbate species are a key factor in the underlying microscopic processes and identify the rotation modes as the ones contributing most to the total measured friction coefficient.
\end{abstract}

\maketitle

% --------------------------
\section{Introduction}
\label{sec:intro}

In every adsorbate system, energy and momentum is transferred between adsorbate and substrate continuously, even when in thermal equilibrium. The substrate can provide the energy required for an adsorbate to explore the potential energy landscape, and can equally absorb the adsorbate's energy, resulting in damping of adsorbate motion. These energy transfer processes influence the adsorbate's behavior strongly and are related to nanoscale friction \cite{krim2012} and vibrational dynamics.\cite{arnolds2011} Activated diffusive motion is dominated by energy exchange, and thus the kinetics of self-assembly processes or surface reactions can be significantly affected. Therefore, understanding the adsorbate--substrate energy transfer, the rate at which it occurs, and the randomization of adsorbate velocities is paramount to the prediction and control of adsorbate dynamical processes and the development of new technologically relevant materials in the fields of surface science and nanoscale materials.

The effects of energy exchange in adsorbate motion can be described quantitatively using the Langevin equation. \cite{zwanzig2001}  Here, key dynamical coordinates are treated explicitly while the influence of all others are treated as being equivalent to that of a heat bath characterized by Gaussian random impulses and a friction parameter. For an adsorbate moving in two-dimensions, the simplest approach is to include only the motion of the center of mass of the diffusing particle. If the center of mass is described by the 2D vector, $\textbf{\textit{r}}$, the simplest form for the equation of motion is
\begin{eqnarray}
  m \ddot{\textbf{\textit{r}}} = - \nabla V(\textbf{\textit{r}}) - m \eta \dot{\textbf{\textit{r}}} + \xi(t),
 \label{eq:langevin}
\end{eqnarray} 
\noindent where the mass, $m$, interacts with the substrate through a potential, $V(\textbf{\textit{r}})$, and is subject to random impulses, $\xi$. The friction coefficient, $\eta$, here is assumed to be position independent and describes the rate of energy transfer between adsorbate and substrate. Energy can be stored in different degrees of freedom of the substrate and adsorbate. Hence, a number of sources contribute to the energy exchange and thus to the friction: electronic terms,\cite{persson1995,altfeder2012} spin,\cite{she2012,wolter2012} phononic degrees of freedom,\cite{StrunzElmer,joostfk,persson1995,persson1999}, and the internal degrees of freedom of the adsorbate.\cite{nijmegendimer,dewijn2009,dewijn2011} The origin and magnitude of the friction depend on the nature of the adsorbate--substrate bond and the electronic properties of the substrate. Phononic friction, for example, acts in both conducting and insulating substrates and, here, the damping of adsorbate motion arises primarily from the emission of substrate phonons. An adsorbate vibration with frequency $\omega_o$ can excite phonons having energies less than $\omega_o$ and, for a Debye spectrum, the number of such states will increases like $\omega_o^3$.\cite{persson1985,persson1999} The friction has a correspondingly strong dependence on the vibrational frequency of the adsorbate. The friction coefficient, $\eta$ is defined by its use in the Langevin equation (Eq.~\ref{eq:langevin}), and as such has inverse time units. It gives a measure of the rate of energy exchange between the various degrees of freedom of the system. In the case of atomic adsorbates energy is typically exchanged 10$^{12}$ times per second, which we denote as ps$^{-1}$. For distinguishable particles in the absence of interactions between adsorbates Eq.~\ref{eq:langevin} gives the well-known result $D\eta=k_B T$, relating the tracer diffusion coefficient, $D$, and the friction in the case of a flat landscape.\cite{zwanzig2001, ala-nissila2002}

Quasi-elastic helium atom scattering provides an experimental measure of the friction by probing adsorbate dynamics. \cite{jardine2002,graham2003} The helium spin-echo (HeSE) technique which we employ here measures the intermediate scattering function, ISF. \cite{jardine2009_ProgSurfSci, squires1978} By comparing measured ISFs with ones derived from Eq.~\ref{eq:langevin} for various values of the friction parameter, $\eta$, a best fit value of $\eta$ may be obtained. The purpose of the present work is to investigate the factors contributing to $\eta$ that arise from the substrate and the internal and external degrees of freedom of the molecule respectively. \cite{dewijn2009,dewijn2011} For that purpose we extend the number of dynamical variables in order to describe each individual atom in the molecule explicitly through interactions between the atoms and the substrate together with internal pairwise and angular forces. The Langevin equation for this extended description is
\begin{eqnarray}
 \begin{aligned} m \ddot{{\vec{r}_i}} =
  &- \nabla \left[V_\mathrm{atom}(\vec{r}_i)+V_\mathrm{molecule}(\vec{r}_i, \{\vec{r}_j\} )\right]  \\
  &- m_i \eta_T \dot{\vec{r}}_i + \xi_{Ti}(t).
  \end{aligned}
\label{eq:fullmd}
\end{eqnarray}
Here, each atom, $i$, in the molecule is subject to random forces, $\xi_{Ti}(t)$, and frictional damping, $\eta_T$, arising from the substrate thermostat, while the forces are derived both from atom--surface and intramolecular atom--atom interactions, $V_\mathrm{atom}(\vec{r}_i)$ and $V_\mathrm{molecule}(\vec{r}_i, \{\vec{r}_j\})$, respectively. Comparison with Eq.~\ref{eq:langevin} and hence with the experimental results is made through a determination of the overall friction, obtained from the trajectory of the center of mass in Eq.~\ref{eq:fullmd}.

\begin{figure}[ht]
 \centering
 \includegraphics[width=85mm]{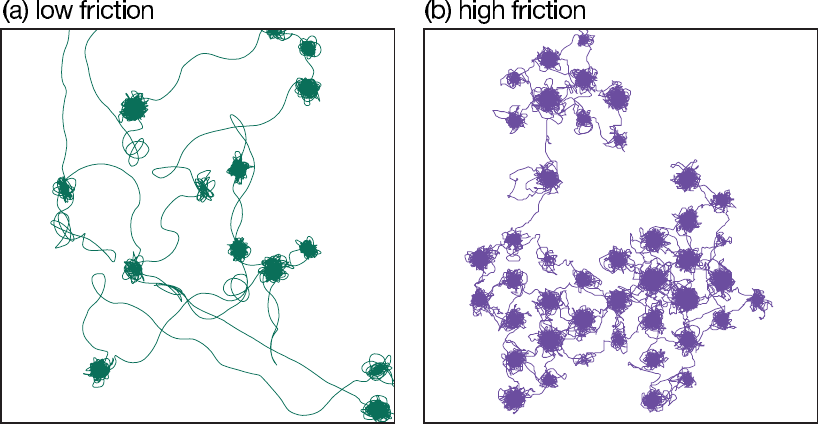}
 \caption{Comparison of center-of-mass trajectories for adsorbates of mass 84~amu experiencing (a) a low friction of $\eta=0.2$~ps$^{-1}$ and (b) a high friction of $\eta=5.0$~ps$^{-1}$ on a substrate of (111) geometry at 160~K. Clear differences in the distribution of jump lengths can be seen, with single jumps between adjacent sites dominating in the high friction regime while longer jumps are observed in the low friction case.}
 \label{fig:lowvshigh}
\end{figure}

Generally speaking, the friction coefficient, $\eta$, determines the distribution of jump lengths for a given adsorbate system. In the case of low friction, the rate of energy exchange between adsorbate and substrate is small and once the diffusing species has acquired sufficient energy to overcome the barrier to diffusion, it ``rollercoasters'' over many barriers, resulting in adsorbate trajectories dominated by long jumps as illustrated in Fig.~\ref{fig:lowvshigh} (a). The adsorbate motion between sites is thus predominantly of a ballistic nature. In the high friction limit, on the other hand, the diffusing particle changes its energy so frequently that it does not move in a direct route between adsorption sites but rather by a random walk. Fig.~\ref{fig:lowvshigh}~(b) shows trajectories dominated by single jumps in a high friction system.

To date, friction coefficients have been determined for a wide range of different adsorbate systems. The diffusion of small adsorbates is typically well understood \cite{jardine2009_ProgSurfSci} and some success has been shown for simple theoretical models of the friction parameter. \cite{persson1999} An investigation of alkali metals on a single crystal surface gave friction coefficients of 0.5~ps$^{-1}$ for Na/Cu(100), \cite{alexandrowicz2006} 0.22~ps$^{-1}$ for K/Cu(100) \cite{hedgeland2009_potassium} and 0.025~ps$^{-1}$ for Cs/Cu(100). \cite{jardine2007} These values for $\eta$ are all in the low friction regime where long jumps dominate the diffusion. Furthermore, a clear trend of decreasing friction with increasing adsorbate mass can be observed. Moving from atomic adsorbates to small molecular species, CO has been investigated on Cu(100),\cite{alexandrowicz2004} Cu(111)\cite{kole2012} and Pt(111),\cite{alexandrowicz2008} giving a friction coefficient of 0.1--0.5~ps$^{-1}$, 0.8~ps$^{-1}$ and 0.7~ps$^{-1}$, respectively, which is on the same order as the friction observed for atomic species. In 
addition, HeSE measurements of propane/Pt(111) have been performed, showing a friction of 0.8$\pm0.2$~ps$^{-1}$. \cite{jardine2008}

Larger molecular adsorbate systems are less well studied and there is a need for new theoretical models to describe these systems. As a consequence, relatively little is known about the friction of such larger adsorbates. Experimental investigation of the benzene/graphite system, \cite{hedgeland2009_benzene} moving in continuous Brownian motion, provided a friction coefficient of $\eta=2.2$~ps$^{-1}$ which is much larger than for any of the smaller systems measured previously. In another helium atom scattering study, W\"{o}ll and coworkers have investigated the frictional damping of the vibrational motion of alkanes on metal surfaces, describing a phononic and an electronic contribution to the friction. \cite{witte1995,witte1997,fuhrmann1998} Their undeconvoluted experimental results were analyzed by Persson and Volokitin who determined a phononic friction coefficient of 2~ps$^{-1}$. \cite{persson1995} More recent work has shown that, in addition to the contributions from the substrate, the internal degrees of freedom of a molecular adsorbate need to be included, giving good agreement between experiment and theory for benzene/graphite.\cite{dewijn2009,dewijn2011} In the present work, we present HeSE measurements of three related aromatic molecules, cyclopentadienyl (Cp), pyrrole and thiophene on Cu(111) and interpret the measured friction coefficients using molecular dynamics simulations including the external as well as internal degrees of freedom of the molecular species.

%
% --------------------------
\section{Methods}

\subsection{The helium spin-echo experiment}

All experiments were performed using the helium spin-echo spectrometer at Cambridge which has been described in detail elsewhere. \cite{jardine2004_Science, fouquet2005} In essence, a spin-polarized $^3$He beam is scattered off the sample surface to measure equilibrium dynamics on the surface. The wave packets of each $^3$He atom are split in time and space by a variable magnetic field into components that scatter from the surface at two different times, $t$ and $t+t_{SE}$. The two components are then recombined in an identical but reversed magnetic field after the sample. For a static surface, both wavepackets scatter identically and are recombined to give the original spin polarization. In the case of surface dynamics occuring during the time difference, $t_{SE}$, (which is of the order of pico- to nanoseconds) a loss of correlation between the surface at $t$ and $t+t_{SE}$ is detected as a reduction in polarization after recombination. During each experiment, we record the increasing loss in correlation on the surface as a function of $t_{SE}$, which typically decays with time as $f(t) = a \exp(-\alpha \cdot t_{SE})+c$ for diffusive motion, where the constant level, $c$, can arise either from scattering from static defects on the surface or from aspects of the motion corresponding to diffusion in a confined space.\cite{jardine2009_ProgSurfSci,paterson2011} To characterize the two-dimensional dynamics of an adsorbate we study the variation of the decay rate, $\alpha$, with scattering momentum transfer parallel to the surface, $\Delta K$, azimuthal direction, temperature and coverage. All experiments reported here were performed using a beam energy between 7.5~meV and 8.5~meV.

A Cu(111) single crystal (Surface Preparation Laboratory, Netherlands) was mounted inside an ultra-high vacuum chamber with a background pressure of $<5~\cdot~10^{-11}$~mbar and cleaned by Ar$^+$ sputtering (800~eV, 10~$\mu$A, 300~K) and annealing to 800~K for 30 seconds. We confirmed the surface cleanliness using the specular helium reflectivity, exceeding $25\%$ at 300~K. The crystal azimuth was aligned using the known diffraction pattern of a CO overlayer. \cite{kole2012}

The dosing process was monitored using the specularly reflected helium signal and the coverage determined from the reflectivity drop. Adsorbed cyclopentadienyl (Cp), C$_5$H$_5$, was obtained by backfilling the chamber with cyclopentadiene (CpH) using the vapor pressure from the liquid sample. The copper surface was kept at 300~K at which temperature CpH dehydrogenates to form Cp on the surface. \cite{hedgeland2011, sun1997} CpH was obtained by distillation of the (CpH)$_2$ dimer and stored at liquid nitrogen temperature to prevent dimerization. Pyrrole, C$_4$H$_4$NH, (Sigma-Aldrich, reagent grade $98\%$) was equally dosed by backfilling the chamber. \cite{lechner2013_pyrrole} A line-of-sight dosing method using a dosing tube in front of the crystal surface was employed for thiophene, C$_4$H$_4$S, (Fluka, puriss. $\geq 99.5\%$). Both pyrrole and thiophene were dosed with the copper crystal kept at 160~K. All adsorption precursors were purified by repeated freeze--thaw cycles in high vacuum before dosing and their purity checked using a quadrupole mass spectrometer. We find that pyrrole and thiophene adsorb reversibly below 200~K, yet the ionically bonded Cp decomposes at temperatures above 500~K.

%
% - - - - - - - - - - - - - 
\subsection{Theoretical approach}

We performed molecular dynamics (MD) simulations to interpret the experimental data and study the microscopic processes governing adsorbate--substrate friction. Simple center-of-mass MD (COM-MD) simulations applying Eq.~\ref{eq:langevin} were used to determine the total friction coefficient from the HeSE data, whereas we employed more complex simulations including the adsorbate species' internal degrees of freedom (IDOF-MD) for a more detailed analysis of the different contributions to the friction coefficient. A comparison of adsorbate trajectories obtained from COM-MD and IDOF-MD is shown in Section~\ref{section:results}.

For our COM-MD, we used an extension of Eq.~\ref{eq:langevin}, including $F_{j,k}$, a pairwise lateral interaction term between molecules $j$ and $k$,
\begin{eqnarray}
  m \ddot{\textbf{\textit{r}}} = - \nabla V(\textbf{\textit{r}}) - m \eta \dot{\textbf{\textit{r}}} + \xi(t) + \sum_{j\not= k} F_{j,k}.
 \label{eq:commd}
\end{eqnarray} 
The potential energy surface (PES), $V(\textbf{\textit{r}})$, is defined by six Fourier coefficients, \cite{alexandrowicz2008,hedgeland2011,lechner2013_pyrrole}
\begin{equation}
  V(\textbf{\textit{r}}) = - \sum_{i,n} A_n \cos(n\textbf{\textit{g}}_i \cdot \textbf{\textit{r}}),
  \label{eq:PES}
\end{equation}
where three vectors $\textbf{\textit{g}}_i$ define the geometry of the hexagonal Cu(111) substrate
\begin{eqnarray}
 \begin{aligned}
 \textbf{\textit{g}}_1&=(\zeta,0), \nonumber \\
 \textbf{\textit{g}}_2&=(\zeta\cos(\pi/3),\zeta\sin(\pi/3)), \nonumber \\
 \textbf{\textit{g}}_3&=(-\zeta\cos(\pi/3),\zeta\sin(\pi/3)).
 \end{aligned}
\end{eqnarray}
$\zeta$ is related to the surface lattice constant, $a$, through $\zeta = 4 \pi/(\sqrt{3} a)$. Using different combinations of first order ($n = 1$) and third order Fourier coefficients ($n = 2$),\footnote{Note that the second order Fourier components for a hexagonal lattice are oriented at 30$^{\circ}$ from the first order ones, hence $n = 2$ corresponds to the third order coefficients.} landscapes for adsorption on top, bridge and hollow sites can be selected. In addition, the shape of the PES can be altered by first scaling it to lie within the range $0\rightarrow1$ on the energy scale, then raising it to a power, $p$, to change the curvature of the PES and finally scaling it back to the original height.\cite{kole2012} Changing $p$ allows flattening or sharpening of the wells at the preferred adsorption sites, making the transition state narrower or wider, respectively. The friction coefficient, $\eta$, the potential coefficients, the scaling power, $p$, and the pairwise lateral interactions are optimized to reproduce the experimental data. At the start of each simulation, randomly allocated particles are given a random kinetic energy in all directions and the system is allowed to relax for typically ca. 320~ps before the core part of the simulation starts. The intermediate scattering function is calculated from the adsorbate trajectories in time steps of 0.16~ps at 25 points in $\Delta \textbf{\textit{K}}$ space. In order to reduce the noise level in the simulations we average over several runs for each simulated curve. The simulated data is analyzed in an identical way as the experimental data to allow direct comparison. A recent study of the dynamics of benzene/graphite showed that repulsive adsorbate interactions can influence the magnitude of the friction.\cite{hedgeland2009_benzene, martinezcasado2007} To rule out such an effect in the current adsorbate systems, COM-MD simulations including lateral interactions were compared with single particle simulations. We found no significant change between the optimized friction coefficient when the interaction term was ignored, suggesting that all systems studied here are in a low coverage regime where inter-adsorbate interactions do not influence the friction.

\begin{figure}[ht]
 \centering
 \includegraphics[width=85mm]{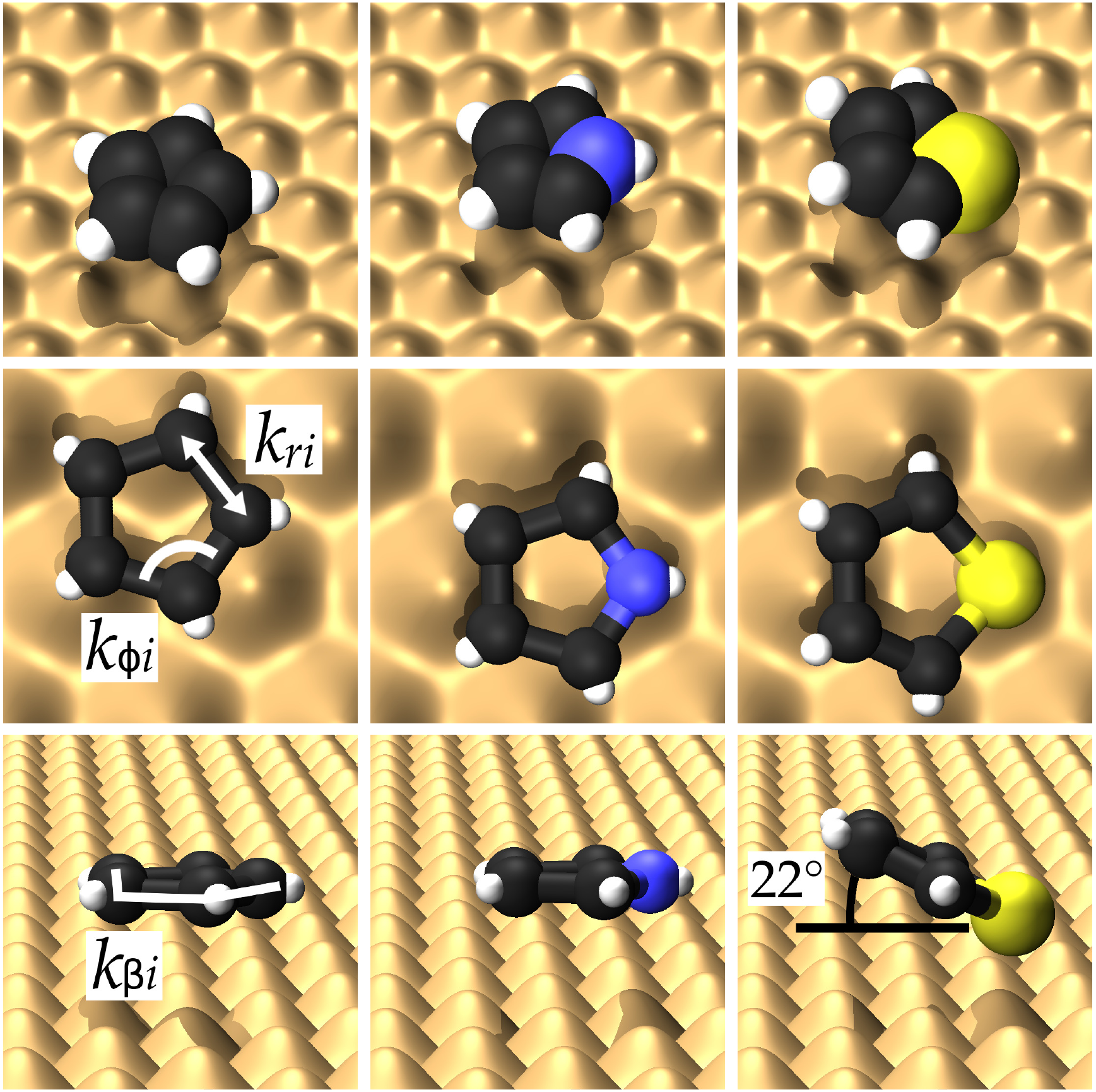}
 \caption{The three molecules (top row), and the bead-spring-like models we use to describe their internal degrees of freedom in a top (center row) and side view (bottom row).  The models include bending $k_{\phi_i}$, stretching $k_{r_i}$, and torsion $k_{\beta_i}$ (departure of four atoms from planar) of the bonds. Cp (left column) is the most symmetric of the three molecules, consisting of five identical CH-complexes, and lies flat on the substrate. Pyrrole (center column) has an NH complex instead of one CH group and also adsorbs flat-lying. Thiophene (right column) has an S atom and adsorbs tilted, not flat, on the substrate.}
 \label{fig:povray}
\end{figure}

To identify the degrees of freedom that play a role in friction, MD simulations based on Eq.~\ref{eq:fullmd}, which include the degrees of freedom of the molecular adsorbates (IDOF-MD), were performed for Cp, pyrrole, and thiophene molecules on a hexagonal Cu(111) substrate. Fig.~\ref{fig:povray} shows the three molecules we consider, and the bead--spring models we use. The total friction arises from the thermal friction from the substrate, $\eta_T$, in addition to the contributions from the molecular degrees of freedom. Specific internal degrees of freedom can be frozen in the simulations, from which their contributions to the friction can be identified.

The value of the frictional damping parameter of the substrate thermostat can be estimated from that for an atomic adsorbate, which is typically between 0.5~ps$^{-1}$ and 1.0~ps$^{-1}$. \cite{alexandrowicz2006} It should be noted that the precise value of $\eta_T$ has relatively little effect on the contributions of the different degrees of freedom to the friction, as demonstrated in Ref. \citenum{dewijn2011}. To allow direct comparison between all of our simulations, we use a fixed value of $\eta_T=1$~ps$^{-1}$. The force field is integrated using a fourth order Runge-Kutta algorithm. Initial conditions at any particular temperature were obtained by allowing the molecule to thermalize at the relevant temperature during an initial period. From simulated trajectories, the change of the center-of-mass velocity over a short time interval was averaged for different values of the velocity. The time interval was chosen long enough so that the correlation in the fast subsystem has time to decay. In the previous work on benzene,\cite{dewijn2011} the correlation time was determined to be around 0.5~ps. The time increment used in the present work to determine the friction coefficient was 0.484~ps=20000~atomic units. The time interval cannot be chosen much longer than this, as on longer time scales, the interaction with the substrate starts to become relevant. The change in velocity over this interval, $\Delta v$, was averaged for different values of the velocity. The total friction coefficient was then extracted by fitting a linear dependence and using the relation $\langle \Delta v\rangle = [\exp(-\eta \Delta t)-1] v $. It should be noted that due to the long trajectories generated by the simulations, the statistical errors in this procedure are small, less than 0.05~ps$^{-1}$, allowing us to compare the results from different simulations. However, there can be larger systematic errors in the final results, due to the uncertainty in the parameters describing the interaction between the atoms and substrate.

The total potential energy of a bond in the molecule consists of contributions from bending, stretching, and torsion. However, due to the low mass of the hydrogen atoms, the energies associated with the bend and stretch vibrational modes involving the C-H and N-H bonds are high compared to the typical energy available at room temperature. Consequently, these bonds will always be in the ground state, and do not participate in the dynamics. \cite{dewijn2009,cvitanovic2010} If an atomistic model is to be used to investigate the internal degrees of freedom, the bonds involving H must therefore be eliminated. Here, we average them out using a mean-field approximation, to obtain a greatly simplified Hamiltonian.\cite{dewijn2009} This leaves five particles for each type of molecule. Each CH, NH, or S is subjected to a substrate potential, as well as a Langevin thermostat, as shown in Eq.~\ref{eq:fullmd}.

Interactions between the atoms are modelled using the Tripos 5.2\cite{clark1989} force field. The internal potential energy of a five-fold ring molecule can be described as a function of the positions, $\vec{r}_i$, bond angles, $\phi_i$ and torsion angles, $\beta_i$,
\begin{eqnarray}
\begin{aligned}
	V_\mathrm{molecule} (\vec{r}_1,\ldots,\vec{r}_5) = \\
	{\textstyle \frac12} \sum_{i=1}^5 k_{r_i}(\|\vec{r}_{(i+1)(\mathrm{mod}~5)}-\vec{r}_{i}\|-r_{0i})^2 \\
+ {\textstyle \frac12} k_{\phi_i} \sum_{i=1}^5 (\phi_i-\phi_{0i})^2 + k_{\beta_i} \sum_{i=1}^5
[1-\cos (2 \beta_i)]~,
\end{aligned}
\label{eq:vmolecule}
\end{eqnarray}
where $k_{r_i}$, $k_{\phi_i}$, and $k_{\beta_i}$ are the force constants for stretching, bending, and torsion, respectively, while $r_{0i}$ and $\phi_{0i}$ are the equilibrium bond lengths and angles. The different degrees of freedom of the molecules are indicated in Fig.~\ref{fig:povray}. The respective force constants and equilibrium bond lengths and angles are given in Table~\ref{table:constants}. The C atoms in Cp are treated somewhat differently from the ones in the other two molecules because Cp is chemisorbed and accepts an electron from the substrate. The atomistic internal dynamics of molecules of this size are typically chaotic, and dominated by the torsion.\cite{dewijn2009,dewijn2011}

\begin{table*}[ht]
\begin{tabular}{llllllllllll}
\hline\hline
& &\multicolumn{2}{c}{stretching}&torsion & &\multicolumn{2}{c}{bending} & &\multicolumn{3}{c}{substrate}\\
%s\hline
& &$r_{0i}$~(\AA{}) & $k_{ri}~(\mathrm{eV}$/\AA$^2$) &$k_{\beta_i}$~(eV) &  & $\phi_{0i}$~(rad) & $k_{\phi i}$~(eV/rad$^2$) & & $z_{0i}$ (\AA) & $V_{0i}$ (meV) &$V_{1i}$ (meV)\\
\hline
Cp &CH-CH &1.395 &121 &0.347 &C-CH-C &$\frac23 \pi$ &6.83 &C-sub. & 0 &-150 & 150\\
\hline
\multirow{3}{*}{pyrrole} &CH-CH &1.395 &121 &0.347 &C-CH-C &$\frac23 \pi$ &10.25 &C-sub. & 0 &-150 & 150\\
&CH-NH &1.346 &113 &0.278 &C-NH-C &$\frac23 \pi$ &14.24 &N-sub. & 0 & -150, 150& 150\\
&			 & & & &C-CH-N &$\frac23 \pi$ &14.24 &			 & & -75, -300 & 75, 300\\
\hline
\multirow{3}{*}{thiophene} &CH-CH &1.395 &121 &0.347 &C-CH-C &$\frac23 \pi$ &10.25 &C-sub. & 0 & 25 & 25\\
				&CH-S  &1.74 &61 &0.0867 &C-S-C  &$\frac{11}{18} \pi$ &11.39 &S-sub. & -0.754 & 56 & 1930\\
					&			 & & & &C-CH-S &$\frac23 \pi$ &21.07 &			 & & &\\
\hline\hline					
\end{tabular}
\caption{Parameter values used in Eqs.~\ref{eq:fullmd}, \ref{eq:vmolecule} and~\ref{eq:vatom} to describe Cp, pyrrole and thiophene on Cu(111) in the IDOF-MD simulations. Bond lengths, force constants and equilibrium angles were taken from Ref.~\citenum{clark1989}, while the potential parameters describing the interaction with the substrate were derived from known experimental and theoretical values. \cite{hedgeland2011, lechner2013_pyrrole, lechner2013, wu1987}}
\label{table:constants}
\end{table*}

Because bulk copper has an fcc lattice, the (111) plane has triangular symmetry. In the IDOF-MD simulations, the substrate is therefore modeled with a triangular two-dimensional substrate potential (using first order Fourier components) and an additional harmonic term in the direction orthogonal to the substrate,
\begin{eqnarray}
V_\mathrm{atom}(\vec{r}) = \frac{2 V_{0i}}{9}\left[2\cos\left(\frac{2 \pi x}{a}\right)\cos\left(\frac{2 \pi y }{a \sqrt{3}}\right)
\right.\nonumber\\
\left.\null + \cos\left(\frac{4 \pi y }{a\sqrt{3}}\right)\right] + V_{1i} \frac{8\pi^2}{9a^2} (z-z_{0i})^2~,
\label{eq:vatom}
\end{eqnarray}
where $a$ is the inter-atomic distance in the triangular lattice of the Cu(111) plane, $a=3.61/\sqrt{2}=2.55$~\AA{}, and $V_{0i}$ and $V_{1i}$ are potential energy scales for the surface corrugation parallel to the surface and curvature in the $z$-direction. The factors of $\frac{2}{9}$ and $\frac{8}{9}\pi^2$ are inserted so that $V_{0i}$ and $V_{1i}$ are of the same order of magnitude, giving equal vibrational frequencies for $|V_{0i}| =|V_{1i}|$.\cite{fuhrmann1998,bruch2004} As we find that small variations of $V_{0i}$ and $V_{1i}$ do not influence the results, we take $|V_{0i}| = |V_{1i}|$ and $z_{0i}=0$ for all atoms except the covalently bonded sulfur atom.

The parameter values used in our simulations are summarized in Table~\ref{table:constants}. We have estimated these parameters based on experimental information and results of density functional theory (DFT) calculations. \cite{hedgeland2011, lechner2013_pyrrole, lechner2013} The sign of the potential corrugation determines the preferred molecular adsorption site, a positive sign giving adsorption on top sites and a negative sign on hollow sites. As experiments show that Cp and pyrrole preferably adsorb on hollow and bridge sites, respectively, giving very similar trajectories moving in channels around top sites, we set $V_{0i}$ for carbon negative for these molecules. The value of -150~meV was estimated from the experimental result for the potential corrugation of a Cp molecule on the Cu(111) substrate,\cite{hedgeland2011} assuming a nearly rigid five-fold ring. For the N atom in pyrrole we have no clear information from experiments and have therefore tested several values for $V_{0i}$, -300~meV, -150~meV, -75~meV, and 150~meV. Thiophene is bound on the top site through the sulfur atom. As the S atom in thiophene is by far the most strongly bound we assume that the single-atom barrier for the S atom for hopping from top to top via bridge sites is similar to the experimentally determined energy barrier of the entire molecule (see Section~\ref{ch:thio}), and thus set $V_{0i} = 56$~meV.\cite{lechner2013} For the other S parameters, we have used values obtained from the vibrational frequency and position of atomic sulfur.\cite{wu1987} The tilted adsorption geometry of thiophene \cite{milligan2001_complete} means that the C atoms are likely to be more weakly bound than in Cp or pyrrole. In order for the C atoms to have the correct distance from the substrate, and for the thiophene molecule to be tilted, the C atoms must have non-zero $V_{1i}$ coefficients. We have thus used a small, non-zero, value for the energy parameter of the C atoms in thiophene that is consistent with the on-top binding of the molecule.\cite{milligan2001_complete} The chosen parameter values lead to an average tilt angle between molecule and substrate of approximately $22^\circ$, comparable to the experimentally determined value of $26\pm5^\circ$ reported for a low coverage of thiophene on Cu(111).\cite{milligan2001_complete}

In order to determine the contribution from various terms, we compare the complete system (all vibrational modes turned on) with systems where some or all internal degrees of freedom are frozen. In particular, we investigate three cases:
\begin{enumerate}[(i)]
\item the full system including contributions from all external and internal degrees of freedom;
\item a system where internal vibrations and frustrated rotations around axes parallel to the surface plane ($x$- and $y$-axes) are frozen, leaving a rigid molecule rotating around the less constrained axis normal to the surface ($z$-axis);
\item a rigid thiophene molecule (i.e. with no internal vibrational modes) which is fully rotating in three dimensions, thus including frustrated rotations around axes parallel to the plane as well as activated rotation\cite{paterson2011} around the $z$-axis.
\end{enumerate}
Note that the external vibrations of the molecules are included in all three models. Simulations of type (ii) give a flat-lying molecule rotating around the geometric center for Cp and pyrrole, and a molecule tilted at $22^\circ$ away from the surface plane rotating around a fixed sulfur atom for thiophene. We should note that because Cp and pyrrole are both flat-lying on the surface, rotation around axes other than the $z$-axis is strongly hindered and therefore simulations of type (iii) do not produce any additional information. In the case of the tilted thiophene molecule, however, frustrated rotations are possible and included in the simulations as they couple to the other dynamical coordinates.

%
% --------------------------
\section{Results}
\label{section:results}

\begin{figure*}[ht]
 \centering
 \includegraphics[width=170mm]{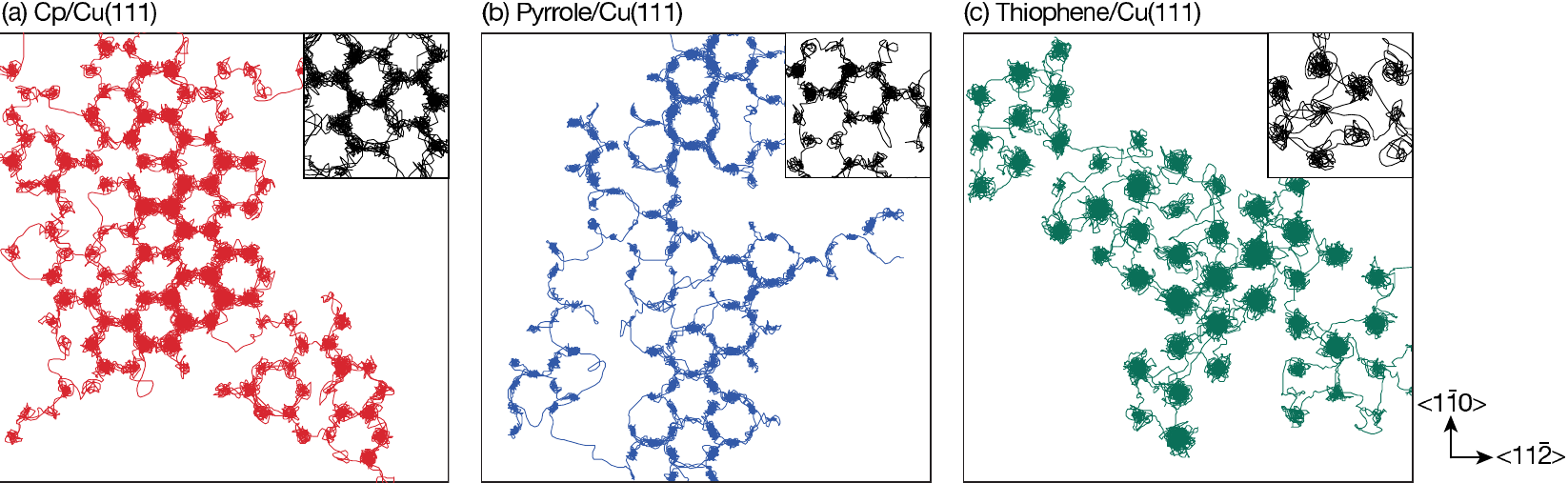}
 \caption{Center-of-mass adsorbate trajectories for (a) Cp at 300~K, (b) pyrrole at 160~K and (c) thiophene at 160~K on a Cu(111) surface as calculated using COM-MD (main panels) and IDOF-MD simulations (insets). COM-MD simulations modeling the experimental data show that all three adsorbates occupy different adsorption sites on the same substrate -- Cp prefers fcc and hcp hollow sites, while pyrrole sits on bridge sites and thiophene bonds to top sites -- yet they all show the clear signature of high friction as the motion is dominated by single jumps between adjacent sites. The respective PESs in the main panels were created using Eq.~\ref{eq:PES} with (a) $n=1$, $p=0.78$ and negative $A_1$, (b) $n=2$, $p=0.6$, negative $A_1$ and positive $A_2$, and (c) $n=1$, $p=1.0$ and positive $A_1$. Comparing the results from COM-MD with the respective trajectories from simulations using Eq.~\ref{eq:fullmd}, shown in the insets, we find good agreement. In the case of pyrrole, we do not reproduce the precise energy landscape in the full simulations, yet the trajectories are comparable. Indeed, deducing the ISFs from simulations on lattices of bridge and hollow sites demonstrates that the two cases give nearly equivalent results in a HeSE measurement.}
 \label{fig:trajectories}
\end{figure*}

The main results of the MD simulations are summarized in Fig.~\ref{fig:trajectories}. The center-of-mass adsorbate trajectories are presented for the three systems Cp, pyrrole and thiophene on Cu(111), where the main panels show the results from COM-MD simulations and the respective insets the corresponding results from IDOF-MD simulations. Whilst we observe qualitative differences between the trajectories for the different adsorbate systems (comparing panels (a), (b) and (c)), the trajectories obtained from the two simulations methods (comparing each panel with its respective inset) show very similar results for each system, adding weight to the value of our comparison.

%
% - - - - - - - - - - - - - 
\subsection{Cyclopentadienyl/Cu(111)}

We have performed HeSE measurements of 0.03~monolayers (ML) of Cp on Cu(111) at 300~K, where we define a ML as one adsorbate particle per substrate atom. A more detailed account of the experiments can be found in Ref.~\citenum{hedgeland2011}. To summarize, momentum transfer dependent data show that Cp hops in single jumps between adjacent fcc and hcp hollow sites. Measuring the motion at variable temperature gives an apparent energy barrier to diffusion of $41\pm1$~meV. COM-MD simulations utilizing a friction coefficient of $\eta = 2.5\pm0.5$~ps$^{-1}$ and a PES that is consistent with DFT calculations, \cite{sacchi2011_cp} provide trajectories with auto-correlation functions that are in excellent agreement with experimental results. By varying $\eta$ in the simulations while keeping all other parameters constant, we find that this friction parameter is at the turnover point from low to high friction, where the jump frequency is maximized.\cite{zwanzig2001} DFT calculations show that Cp is an ionically bonded adsorbate, yet the frictional coupling strength is very similar to that observed for the physisorbed benzene/graphite system,\cite{hedgeland2009_benzene} implying that adsorption strength does not influence the lateral friction. Furthermore, a flat potential energy landscape was found for benzene/graphite while Cp/Cu(111) shows activated behavior, illustrated in the trajectories obtained from MD simulations, shown in Fig.~\ref{fig:trajectories}~(a).

Trajectories obtained from the IDOF-MD simulations of Cp are shown as an inset to Fig.~\ref{fig:trajectories} (a) and look similar to those from the COM-MD simulations. The IDOF-MD simulations for the full system -- model (i) -- give a friction coefficient of 3.7~ps$^{-1}$, indicating a significant contribution of the molecular degrees of freedom to the total friction compared to the friction arising purely from the substrate thermostat, which we have set to $\eta_T = 1$~ps$^{-1}$. The fact that the simulated value is somewhat higher than the experimental friction is most likely due to the limitations of the pairwise description between the atoms in the chemisorbed molecule and the substrate, compounded by the complications in estimating the parameters for the interaction and $\eta_T$. Both experimental and theoretical values describe similar adsorbate trajectories, however, dominated by single jumps between adjacent sites, and thus give good overall agreement. Repeating the IDOF-MD simulations without vibrational dynamics -- model (ii) -- we find a nearly identical friction coefficient as for the full system, shown in Table~\ref{table:exptvstheory}. From this, we conclude that the rotational dynamics of the molecule around the $z$-axis dominate the friction of Cp.

%
% - - - - - - - - - - - - - 
\subsection{Pyrrole/Cu(111)}

Pyrrole is found to adsorb in a flat-lying geometry predominantly through van der Waals bonding, centered over bridge sites on Cu(111). HeSE measurements of 0.033~ML pyrrole/Cu(111) at 160~K, which have been reported elsewhere,\cite{lechner2013_pyrrole} show hopping motion dominated by single jumps, similar to the behavior of Cp on the same substrate. In addition, strong lateral interactions are evident in the data, manifesting in a strong de Gennes narrowing feature,\cite{degennes1959} indicating that the adsorbates repel each other. Temperature dependent measurements provide an apparent activation energy of $50\pm3$~meV. MD simulations give a friction coefficient of $2.0\pm0.4$~ps$^{-1}$ which is again on the same order as that of benzene/graphite \cite{hedgeland2009_benzene} and Cp/Cu(111).\cite{hedgeland2011}

The IDOF-MD results for the friction of pyrrole depend somewhat on the parameter values used for $V_{0i}$ and $V_{1i}$ and vary between 2.6~ps$^{-1}$ and 3.9~ps$^{-1}$. The highest value was obtained for $V_{0i}=-150$~meV and $V_{1i}=150$~meV, i.e. when the interaction parameters for the N atom were identical to those for the C atoms. With increasing asymmetry, the friction value decreases, both for increased and decreased coupling. As was the case with Cp, the friction values with -- model (i) -- and without vibrational dynamics -- model (ii) -- are nearly identical and thus the rotational motion of the molecule about the $z$-axis must dominate the friction.

%
% - - - - - - - - - - - - - 
\subsection{Thiophene/Cu(111)}
\label{ch:thio}

A comparison of the atomic adsorbate systems summarized in section \ref{sec:intro} with benzene/graphite,\cite{hedgeland2009_benzene} Cp/Cu(111),\cite{hedgeland2011} and pyrrole/Cu(111) \cite{lechner2013_pyrrole} suggests that the frictional coupling is stronger in large molecular systems, irrespective of bonding mechanism and potential energy landscape corrugation. To investigate the effect of adsorption geometry on the friction, we have now extended our study to a third five-membered aromatic adsorbate, thiophene, on Cu(111).

\begin{figure}[ht]
 \centering
 \includegraphics[width=85mm]{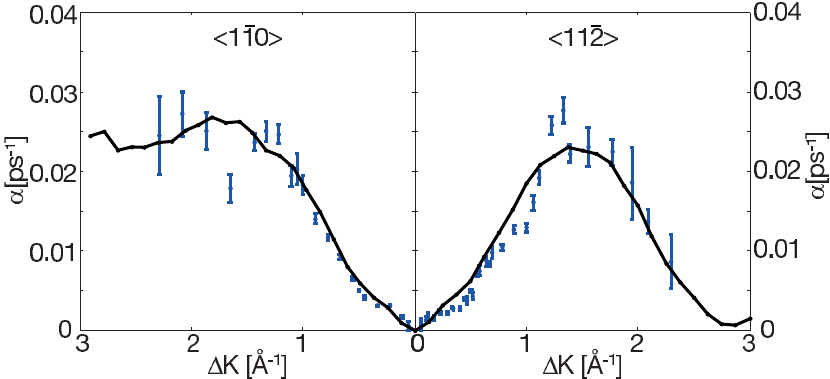}
 \caption{HeSE measurements of the dynamics of 0.015~ML thiophene/Cu(111) at 160~K along the two main crystal directions $<$1$\bar{1}$0$>$ (left panel) and $<$11$\bar{2}$$>$ (right panel). Experimental data points, shown in blue, are compared with results from simple center-of-mass molecular dynamics simulations (solid line) providing a friction coefficient of $\eta = 5\pm2$~ps$^{-1}$.}
 \label{fig:thiophene}
\end{figure}

Fig.~\ref{fig:thiophene} shows HeSE measurements of 0.015~ML thiophene/Cu(111) at 160~K, recorded along two azimuthal directions. The sinusoidal shape of the experimental curves follows that predicted by analytical models for hopping on a simple Bravais lattice,\cite{chudley1961} confirming Milligan et al.'s findings that thiophene adsorbs preferably on top sites.\cite{milligan2001_complete} Small peak and dip deviations from a sinusoid can be found at low values of $\Delta \textit{\textbf{K}}$ which indicate repulsive inter-adsorbate interactions, allowing a coverage determination from the position of the dip.\cite{degennes1959} Temperature dependent measurements give an Arrhenius barrier to diffusion of $E_a = 59\pm2$~meV.

We interpret the experimental data with COM-MD simulations. By optimizing the PES and friction coefficient we obtain excellent agreement with the $\alpha(\Delta \textbf{\textit{K}})$ curve, shown as solid lines in Fig.~\ref{fig:thiophene}, and the Arrhenius activation energy. The best fit to the data was obtained with a friction coefficient of $\eta=5\pm2$~ps$^{-1}$, which is exceptionally high.

The IDOF-MD simulations including all degrees of freedom -- model (i) -- show a friction value of 3.2~ps$^{-1}$ for thiophene/Cu(111), which is within the error bars of the experimental value. Simulations of a rigid thiophene molecule tilted at a constant angle -- model (ii) -- produce a similar, slightly higher, value for the friction. However, when the molecule has more degrees of freedom and is allowed to rotate in 3 dimensions -- model (iii) -- the friction is lowered to 2.9~ps$^{-1}$. This indicates that, besides the rotation around the axis normal to the substrate, other dynamics, such as frustrated rotations, also play a role.

%
% --------------------------
\section{Discussion}

Comparing the previous results for benzene on graphite \cite{hedgeland2009_benzene, dewijn2009, dewijn2011} with our findings for Cp, pyrrole and thiophene on Cu(111), as summarized in Table~\ref{table:exptvstheory} for both theory and experiment, we can deconstruct the influence of microscopic parameters on the friction coefficient. From an experimental point of view, we find that the friction is fairly insensitive to the corrugation of the potential energy landscape, as the friction coefficient for benzene on a flat graphite landscape is very similar to that for Cp and pyrrole on the corrugated Cu(111) surface. Similarly, we do not observe any trends with the adsorbate--substrate bond strength when comparing the ionically bonded Cp with the physisorbed benzene and pyrrole. Finally, by extending the existing experiments to the tilted thiophene/Cu(111) system, we can also investigate the role of adsorption geometry. Somewhat surprisingly, the friction coefficient obtained for thiophene is the largest amongst all four systems studied, even when taking into account the rather large error bars. Our experimental results thus suggest that the addition of frustrated rotational degrees of freedom for a molecule tilted away from the surface increases the friction, although this effect is not reflected in the present simulations.

\begin{table}[ht]
 \centering
 \begin{tabular}{l l lll}
 \hline
 \textbf{Adsorbate system} 	&\textbf{$\eta_{\text{COM-MD}}$ [ps$^{-1}$]} 	&\multicolumn{3}{l}{\textbf{$\eta_{\text{IDOF-MD}}$ [ps$^{-1}$]}} \\
 &&(i)&(ii)&(iii)\\
 \hline
 Benzene/graphite		&2.2\cite{hedgeland2009_benzene}		& 1.9~\cite{dewijn2011} &1.8~\cite{dewijn2011}&-\\
 Cp/Cu(111)			&$2.5\pm0.5$\cite{hedgeland2011}		& 3.7 &3.7&-\\
 Pyrrole/Cu(111)		&$2.0\pm0.4$\cite{lechner2013_pyrrole}			& 2.6-3.9 &2.6-3.9&-\\
 Thiophene/Cu(111)		&5$\pm$2					& 3.2 & 3.4 &2.9\\
 \hline
 \end{tabular}
 \caption{Comparison of friction coefficients for a range of aromatic adsorbate systems. The contributions of the different molecular degrees of freedom to the friction are determined by modeling (i) the full system including contributions from all degrees of freedom, (ii) rigid molecules with frozen internal modes, leaving rotation around the $z$-axis as the only included degree of freedom, and (iii) in the case of thiophene a system including the contributions from frustrated rotations normal to the surface plane and rotation around the $z$-axis. The statistical errors in the IDOF-MD values are small, less than 0.05~ps$^{-1}$, allowing the comparison of the results from models (i), (ii) and (iii). However, there can be larger systematic errors in the final results, due to the uncertainty in the parameters of the potential-energy landscape and $\eta_T$. All theoretical results shown here are for a thermal friction of $\eta_T=1$~ps$^{-1}$. These uncertainties could account for the slight discrepancy between some experimental and theoretical values, yet do not affect our conclusions regarding the contributions of the molecular degrees of freedom.
}
\label{table:exptvstheory}
\end{table}

Previous work showed that for benzene/graphite the dominating degrees of freedom are rotation and torsion,\cite{dewijn2009,dewijn2011} whilst here we find rotation around the axis normal to the substrate clearly dominates the friction in all three five-membered molecules on Cu(111). From Cp to thiophene, the molecules become increasingly asymmetric. Pyrrole, unlike Cp, has one atom that is different from the others. By varying the coupling strengths for the N atom in the IDOF-MD simulations of pyrrole, we have investigated the effects of this asymmetry on the friction. The friction is at a maximum if the N and C have the same coupling strength to the substrate. As all contributions to the friction are mediated through the coupling between the individual atoms and the substrate, \cite{dewijn2011} a reduction in friction with weaker coupling is to be expected. However, an increase of a factor of two in the coupling strength of the N atoms also reduces the friction by a third; as rotation, which is the main channel of dissipation, is partially blocked. Because of the (approximate) five-fold symmetry of the molecule and triangular symmetry of the lattice, a fully symmetric molecule feels only a small energy barrier against rotation around the $z$-axis. When the N atom couples more strongly the barrier increases and rotation becomes more difficult.

In comparison, further asymmetry is added in the thiophene/Cu(111) system by the fact that the molecule bonds to the substrate covalently through the sulfur atom with the ring tilting away from horizontal. By comparing the friction obtained for the full system in model (i) with cases where different modes are turned off in the IDOF-MD simulations, we can deconstruct the contributions of the different modes. In case (iii) only the internal vibrations are switched off, while in case (ii) the frustrated rotation modes are missing as well. From the friction values given in Table~\ref{table:exptvstheory}, we find that frustrated rotations give an important contribution to the friction in an adsorbate that is not flat-lying. We further note that in the case of thiophene the internal vibrational degrees of freedom decrease the friction somewhat, going from model (ii) to (i). The mechanisms causing this small decrease are complex and most likely due to a change in the coupling between the different modes. Overall, it must be concluded that several modes contribute significantly, and in either sense, to the frictional forces experienced by thiophene.

The present work demonstrates that MD simulations taking into account the internal degrees of freedom of a molecular adsorbate species can be a valuable tool for the analysis of HeSE experiments. In addition to the investigation of friction coefficients and center-of-mass adsorbate trajectories described in the present work, it is possible to deduce ISFs from IDOF-MD simulations for a more direct comparison with HeSE data, as is routinely done for COM-MD simulations.\cite{hedgeland2009_benzene,kole2012,paterson2011} In general, the number of free parameters in IDOF-MD simulations is likely too high to allow a reliable analysis of new HeSE measurements. Once additional information about the potential energy landscape has been obtained (e.g. by COM-MD or DFT calculations), however, we believe IDOF-MD simulations can be a valuable tool to test a PES or to investigate the different contributions to the friction in more detail.

%
% --------------------------
\section{Summary}

In conclusion, we have shown the importance of internal degrees of freedom in the friction of molecular adsorbates. By comparing four systems that span a range of different adsorption geometries, bond strengths and energy landscape corrugations, we have demonstrated that the influence of these three parameters is relatively unimportant in the friction for these systems, while the increase in friction compared to atomic adsorbate systems is large. IDOF-MD simulations show that the main contribution to the increase in friction of the molecular systems can be attributed to frustrated rotation around axes parallel to the surface plane and activated rotation around the $z$-axis.

%
% --------------------------
\begin{acknowledgments}
BAJL wishes to thank the Austrian Academy of Sciences and the EPSRC for funding, ASdW acknowledges financial support by an Unga Forskare grant from the Swedish Research Council, APJ is grateful for a Royal Society University Fellowship, and BJH thanks the US National Science Foundation (CHE1124879) for support. Furthermore, financial support by the EPSRC (EP/E0049621) is gratefully acknowledged by all Cambridge authors.
\end{acknowledgments}

%
% --------------------------
%

\end{document}